\documentstyle{l-aa} 

\begin{document} 
\thesaurus{01.01.1, 03.10.1, 07.01.1, 07.13.1, 17.03.1, 19.04.1}       
 
\title{Acceleration of UHE Cosmic Ray Particles at Relativistic   
Jets in Extragalactic Radio Sources}   
    
\author{M. Ostrowski}

\institute{ Obserwatorium Astronomiczne, Uniwersytet    
 Jagiello\'nski, ul.Orla 171, 30-244 Krak\'ow, Poland}    
\offprints{M. Ostrowski (mio@oa.uj.edu.pl)}       
\date{Received..; accepted ..;} 
 
\maketitle 
 
\begin{abstract}{A mechanism of ultra-high energy cosmic ray acceleration in 
extragalactic radio sources, at the interface between the {\it 
relativistic} jet and the surrounding medium, is discussed as a 
supplement to the shock acceleration in `hot spots'. Due to crossing the 
tangential discontinuity of the velocity the particle can gain an amount 
of energy comparable to the energy gain at the shock crossing. However, 
the spectrum of particles accelerated at the jet side boundary is 
expected to be much flatter than the one formed at the shock. Due to 
this fact, particles accelerated at the boundary can dominate the 
overall spectrum at highest energies. In conditions characteristic to 
extragalactic jets' terminal shocks, the mechanism naturally provides 
the particles with $E \sim 10^{20}$ eV and complies with the efficiency 
requirements. The spectrum formation near the cut-off energy due to 
action of both the shock acceleration and the tangential discontinuity 
acceleration is modelled with the Monte Carlo particle simulations. It 
confirms that the upper energy limit can surpass the shock acceleration 
estimate.} 
    
\end{abstract} 

\section{INTRODUCTION}    
    
Jet-like outflows are observed in a number of astrophysical   
environments, starting from young stars embedded in their parent   
molecular clouds, up to active extragalactic objects. In the later   
case, a number of interesting observational phenomena are noted over   
orders of magnitude of linear sizes. In particular, at the smallest   
mili-arc-second scales, one often observes relativistic jet velocities,   
with flow Lorentz factors $\gamma_u$ reaching the values above $10$ (cf.  
Ghisellini et al. 1996). At larger scales, velocity measurements are 
more difficult, but without entrainment of large amount of matter near  
the active galactic nuclear source the jet flow velocity must be also  
relativistic. A possible loading of a jet with matter is expected to be  
appended by a substantial amount of turbulence (Henriksen 1987) and  
related jet kinetic energy dissipation. However, in the FR~II radio  
sources, there are often observed jets efficiently transporting energy  
to the far-away hot spots and any jet breaking mechanism can not act too  
effectively near the central core. Also, the existing hydrodynamical 
simulations of relativistic jets show for possibility of extended stable 
jet structures (Marti et al. 1995, 1997; G\'omez et al. 1995). Another 
argument suggesting the relativistic jet speed at all scales, may be 
based on the visible asymmetry of jets with respect to the nuclear 
source, if one believes the effect is caused by the high velocity of the 
essentially bi-symmetric outflow (cf. Bridle et al. 1994). Let us also 
note that the Meisenheimer et al. (1989) modelling of the shock 
acceleration process at extragalactic radio-source hot spots yields `the 
best-guess' jet velocities in the range ($0.1$, $0.6$) 
    
The relativistic movement of the jet leads to shock wave formation in places 
where an obstacle or perturbation of the flow creates a sudden velocity   
jump. For jets loaded with a cold plasma the highly oblique conical   
shocks are formed within the jet tube. These shocks can have a  
non-relativistic character, involving the velocity jump perpendicular to   
the shock surface much smaller than the overall jet velocity $U \sim c$.   
They lead to a limited kinetic energy dissipation and are usually   
claimed to be responsible for forming the so called `knots' along the   
jet. A much more powerful shock is formed at the final working surface of 
the jet. There, a substantial fraction of the jet energy is transferred   
into heating the jet's plasma, generating strong turbulence, boosting 
magnetic fields within the turbulent volume, and finally accelerating 
electrons and nuclei to cosmic ray energies. Rachen \& Biermann (1993) 
considered the process of particle acceleration to ultra-high energies 
(UHE) at such shocks. They show that given the favourable conditions the 
UHE particles up to $\sim 10^{20}$ eV can be formed. Then Rachen et al. 
(1993) show that assumption of UHE particle acceleration in 
extragalactic powerful radio sources is compatible with the current 
measurements of cosmic ray abundances and spectra at energies above 
$10^{17}$ eV. Additionally, the arrival directions of cosmic ray 
particles observed above $10$ EeV are correlated with the local galactic 
supercluster structure (Stanev et al. 1995; see, also, Medina Tanco et 
al. 1996, Sigl 1996, Sigl et al. 1995, 1996, Hayashida et al. 1996, 
Elbert \& Sommers 1995, Geddes et al. 1996 and Medina Tanco 1998). An 
alternative model involving the several-Mpc-scale non-relativistic 
shocks in galaxy clusters is proposed by Kang et al. (1996; see also 
Kang et al. 1997). 
 
As noted by us (Ostrowski 1990; henceforth Paper~I) a tangential 
discontinuity of the velocity field can also provide an efficient cosmic 
ray acceleration site if the considered velocity difference $U$ is 
relativistic and the sufficient amount of turbulence on both its' sides 
is present. The problem was extensively discussed in the early eighties 
by Berezhko with collaborators (see the review by Berezhko 1990) and in 
the diffusive limit by Earl et al. (1988) and Jokipii et al. (1989). In 
the present paper we consider the process of ultra high energy cosmic 
ray acceleration in relativistic jets including the possibility of such 
boundary layer acceleration. As the considerations of Rachen \& Biermann 
(1993) treat the acceleration process at relativistic shock in a 
somewhat simplified way (see, also Sigl et al. 1995), in the first part 
of the next section (section 2.1) we review this process in some detail 
in order to understand the inter-relations between the conditions 
existing near the shock, the accelerated particle spectrum and the 
particle's upper energy limit. Then, in section (2.2), we present a short 
description of the basic physical model for the considered acceleration 
process acting at the jet boundary. We show (section 2.3) that in the 
conditions characteristic for relativistic jets in extragalactic radio 
sources, particles with energies above $10^{20}$ eV can be produced in 
this process without extreme parameter fitting. The required efficiency 
is discussed in section (2.4). We confirm the estimates presented 
previously for the shock acceleration, showing that the UHE particle flux 
observed at the Earth can be reproduced as a result of acceleration 
processes in jets of nearby powerful radio sources. In section 3 we 
discuss the problem of the particles' spectrum. With the use of Monte Carlo 
simulations, we consider the action of both processes acting near the 
terminal shock in a relativistic jet. Modification of the spectrum due to 
varying boundary conditions and jet velocity is discussed for the case 
of ($e^-$, $p$) jets expected to occur in the powerful FRII radio 
sources (cf. Celotti \& Fabian 1993). The derived particle's upper energy 
limits are above the shock acceleration estimates and the spectrum 
modification at highest energies can resemble the observed above 10 EeV 
`ankle' structure. A short summary and final remarks are presented in 
section 4. A preliminary report about this work was presented in 
Ostrowski (1993b, 1996). 
    
For the discussion that follows, we consider the jet propagating with    
the relativistic velocity, $U \sim c$. We use $c$ = 1 units.    
    
\section{ACCELERATION PROCESSES IN RELATIVISTIC JETS}    
    
The present considerations attempt to extend the discussion of particle   
acceleration at shock waves formed in the end points of jets by including  
an additional acceleration process acting at the jet boundary layer.  
For particles with UHE energies both the shock transition as well as   
the velocity transition layer between the jet and the surrounding   
medium (`cocoon') can be approximated as surfaces of discontinuous   
velocity change\footnote{One may also note that in the perfect fluid   
simulations of relativistic jets by Marti et al. (1995, 1997) the numerical  
viscosity inherent to such approach does not lead to generation of an   
extended shear layer.}. Basing on such approximation we compare the   
acceleration at the non-compressive tangential discontinuity at the jet   
side boundary and the compressive terminal shock discontinuity.   
    
\subsection{Shock acceleration in a hot spot}    
    
A review of the problems related to energetic particle acceleration at  
relativistic shock waves is presented by Ostrowski (1996) and Kirk  
(1997). In the present section we summarize some most important findings  
in this subject. The main difficulty in considering cosmic ray  
acceleration at relativistic shocks arises from the substantial particle  
anisotropies involved. A consistent approach to the first  
order Fermi acceleration at such a shock propagating along the  
background magnetic field (`parallel shock') was conceived by Kirk \&  
Schneider (1987) who obtained solutions to a kinetic equation of the 
Fokker--Planck type with a pitch-angle diffusion scattering term. More 
general conditions at the parallel shock were considered by Heavens \& 
Drury (1988), who took into consideration the fluid dynamics of 
relativistic shock waves. For a shock propagating in the cold (e, p) 
plasma a trend was revealed to make the accelerated particle spectrum 
slightly flatter ($\sigma \approx 3.7$) for the shock velocity, $U$, 
growing up to roughly $0.5 c$, and, then, increasing the inclination 
with further growth of $U$ up to $\sigma \approx 4.2$ at the highest 
considered velocity $u = 0.98$. However, the resulting varying spectral 
index could still be reasonably approximated with the non-relativistic 
expression $\sigma = 3R/(R-1)$, where $R$ is the shock compression 
ratio. They also noted that the spectrum inclination depends on the 
perturbations' spectrum near the shock, in contrast to the 
non-relativistic case. A qualitatively new possibility was revealed by 
Kirk \& Heavens (1989) who considered the acceleration process in shocks 
with magnetic fields oblique to the shock normal (see also Ballard \& 
Heavens 1991 and Ostrowski 1991). They demonstrated, again in contrast 
to the non-relativistic results, that such shocks led to flatter spectra 
than do the parallel ones, with $\sigma \approx 3.0$ for the 
(subluminal) quasi-perpendicular shocks. Their work relied on the 
assumption of adiabatic invariant $p_\perp^2 /B$ conservation for 
particles interacting with the shock and, thus, was limited to only 
slightly perturbed background magnetic fields. A different approach to 
particle acceleration was presented by Begelman \& Kirk (1990), who 
noted that in relativistic shocks most field configurations lead to 
super-luminal conditions. Then particles can be energized in a single 
shock transmission only, accompanied with a limited energy gain, but the 
acceleration in relativistic conditions is more efficient than that 
predicted by a simple adiabatic theory. The acceleration process in the 
presence of finite amplitude perturbations of the magnetic field was 
considered by Ostrowski (1991; 1993a), Ballard \& Heavens (1992) and 
Bednarz \& Ostrowski (1996). The considerations involved the Monte Carlo 
particle simulations for shocks with oblique perturbed magnetic fields. 
It was noted that the spectral index was not a monotonic function of the 
perturbation amplitude, enabling for the steeper spectra at intermediate 
perturbation amplitudes than those for the limits of small and large 
amplitudes. It has also been revealed that the conditions leading to 
very flat spectra involve an energetic particle density jump at the 
shock and probably lead to instability. The acceleration process in the 
case of a perpendicular shock shows a transition between the compressive 
acceleration described by Begelman \& Kirk (1990) and, at larger 
perturbations, the regime allowing for formation of a wide range 
power-law spectrum. As a conclusion of this short review one should note 
that the present theory is unable to predict the spectral index of 
particles accelerated at the relativistic shock wave, e.g., the possible 
range of indices arising in computations for the sub-luminal shocks 
propagating in the cold ($e$, $p$) plasma is $3.0 < \sigma < 4.5$. 
   
To date, there was somewhat superficial information on the acceleration  
time scales, $T_{acc}$, in relativistic shocks as the applied approaches 
often neglected or underestimated a significant factor controlling the  
acceleration process -- the particle anisotropy. The realistic particle  
distributions are considered in Bednarz \& Ostrowski (1996; see also  
Ellison et al. 1990 and Naito \& Takahara 1995 for specific cases) who  
considered shocks with oblique, sub- and super-luminal magnetic field  
configurations and with finite amplitude perturbations, $\delta B$. At  
parallel shocks, $T_{acc}$ diminishes with increasing perturbation 
amplitude and the shock velocity $U_1$. A new feature discovered in 
oblique shocks is that due to the cross-field diffusion $T_{acc}$ can  
change with $\delta B$ in a non-monotonic way. The acceleration process  
at the super-luminal shock leading to the power-law spectrum is possible 
only in the presence of a large amplitude turbulence. Then, in contrast to 
the quasi-parallel shocks, $T_{acc}$ increases with the increasing wave  
amplitude. For {\it mildly} relativistic shocks, in some magnetic field  
configurations one discovers a possibility to have extremely short 
acceleration time scales, comparable, or even smaller than the particle  
gyroperiod in the magnetic field upstream of the shock. It is also noted 
that there exist a coupling between the acceleration time scale and the  
resulting particle spectral index. Again, the above variety of different  
results illustrates the difficulty in providing an accurate acceleration  
time scale estimate without a detailed knowledge of the conditions in 
the shock. 
\begin{figure} 
\vspace {10cm} 
\includegraphics{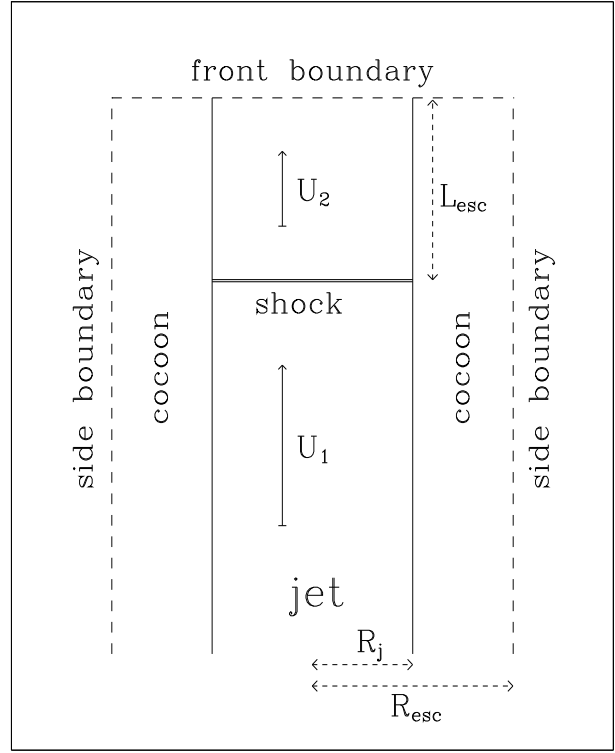}    
\caption[1]{A schematic representation of the terminal shock   
neighbourhood. The velocities and distances used in the text are   
indicated.}   
\end{figure}

A discussion of UHE particle acceleration at mildly relativistic shocks 
formed at the powerful radio source hot spots was presented by Rachen \& 
Bierman (1993) and Sigl et al. (1995). The partly qualitative 
considerations show a potential difficulty in accelerating particles to 
the highest required energies. Besides the difficulties with the 
acceleration time scale there are even more severe constraints for the 
particle energy due to the boundary conditions: the finite perpendicular 
extent of the jet and the finite extent of the shock's downstream region 
situated within the radio source hot spot. The considerations of the 
time dependent acceleration at shocks described above (Bednarz \& 
Ostrowski 1996) allow for a very rapid acceleration only in some 
particular conditions. One should also be aware of the another problem. 
As discussed above, due to an anisotropic particle distribution at the 
shock, the different physical factors acting near it can substantially 
modify the particle energy distribution. Thus, let us stress again, the 
theory is not able to predict every particular spectral index for the 
accelerated particles (cf. Ostrowski 1994, 1996) or any particular 
acceleration time scale and any attempt to compare such existing 
predictions to the observations bears a substantial degree of 
arbitrariness. 
   
\subsection{Acceleration process at the jet side boundary}  
    
A tangential discontinuity of the velocity field (or a shear layer) 
occurring at the jet side boundary can be an efficient cosmic ray 
acceleration site if the considered velocity difference, $U$, is 
relativistic and the sufficient amount of turbulence on its both sides 
is present \footnote{For simplicity, we consider the magnetic field 
perturbations static in the local plasma rest frame and thus we neglect 
any additional acceleration due to the second-order Fermi process.} 
(Paper~I, Ostrowski 1997). If near the jet boundary particles exist with 
gyroradii (or mean free paths normal to the boundary) comparable to the 
actual thickness of the shear-layer interface, the acceleration process 
can be very rapid. One may note, that the particles with energies $> 1$ 
EeV, of interest here, could satisfy the last condition naturally. Any 
high energy particle crossing the boundary from, say, region I (within 
the jet) to region II (off the jet), changes its energy, $E$, according 
to the respective Lorentz transformation. It can gain or loose energy. 
In the case of a uniform magnetic field in region II, the successive 
transformation at the next boundary crossing, II $\rightarrow$ I, 
changes the particle energy back to the original value. However, in the 
presence of perturbations acting at particle orbits between the 
successive boundary crossings there is a positive mean energy change: 
    
$$<\Delta E>~=\, \rho_E \, (\gamma_u-1) \, E \hskip 2cm , \eqno(1)$$  
    
\noindent    
where $\gamma_u \equiv (1-U^2)^{-1/2}$ is the flow Lorentz factor. The 
numerical factor $\rho_E$ depends on particle anisotropy at 
discontinuity. It increases with the growing field perturbations' 
amplitude but slowly decreases with growing flow velocity. Particle 
simulations described in Paper~I give values for $\rho_E$ within the 
strong scattering limit as a substantial fraction of unity. During the 
acceleration process, particle scattering is accompanied by the jet's 
momentum transfer into the medium surrounding it. On average, a single 
particle with momentum $p$ transports the following amount of momentum 
across the jet's boundary: 
    
$$<\Delta p>\, =\, <\Delta p_z>\, =\,  \rho_p \, U \, p \hskip 2cm ,    
\eqno(2)$$  
    
\noindent    
where the value of $p$ is given as the one after transmission and the   
$z$-axis of the reference frame is chosen along the flow velocity. The 
numerical factor $\rho_p$ depends on the scattering conditions near the 
discontinuity and it can reach values also being a substantial fraction 
of unity. As a result, there exists a drag force {\it per unit surface} 
of the jet boundary acting on the medium along the jet, of a magnitude 
order same as the accelerated particles' energy density. Independent of 
the exact value of $\rho_E$, the acceleration process can proceed very 
fast in the case of the mean magnetic field being parallel to the 
boundary (cf. Coleman \& Bicknell 1988, Fedorenko \& Courvoisier 1996) 
because particles can be removed at a distance from the accelerating 
interface only in the inefficient process of cross-field diffusion. One 
may note that in the case of a non-relativistic velocity jump, $U \ll 
c$, the acceleration process becomes of the second-order in $U/c$ and a 
rather slow one. 
    
\subsection{The upper energy limit for the tangential discontinuity    
acceleration}    
    
For acceleration at the considered tangential discontinuity the  
acceleration time scale $T_{acc}$ -- as long as the flow is mildly 
relativistic -- can be approximately determined by the mean time between 
boundary crossings, $<\Delta t>$, and the mean energy gain at the 
crossing, $<\Delta E>$ (Eq.~1; cf. Bednarz \& Ostrowski (1996) for the 
case of non-vanishing correlations between $\Delta t$ and $\Delta E$, 
when $\Delta E \sim E$). Within a simple diffusive model in Paper~I, the 
time $T_{acc}$ is expressed with the use of $\lambda / c$, where 
$\lambda$ is the particle mean free path. $T_{acc}$ depends on a number 
of physical factors, including jet velocity and magnetic field structure 
determining parameters of particle wandering at the acceleration region 
and the mean energy gain at individual boundary crossing. Thus we 
express the acceleration time scale as 
    
$$T_{acc} = \alpha \, \lambda \, / \, c \qquad , \eqno(3)$$  
\begin{figure} 
\vspace{18cm} 
\includegraphics{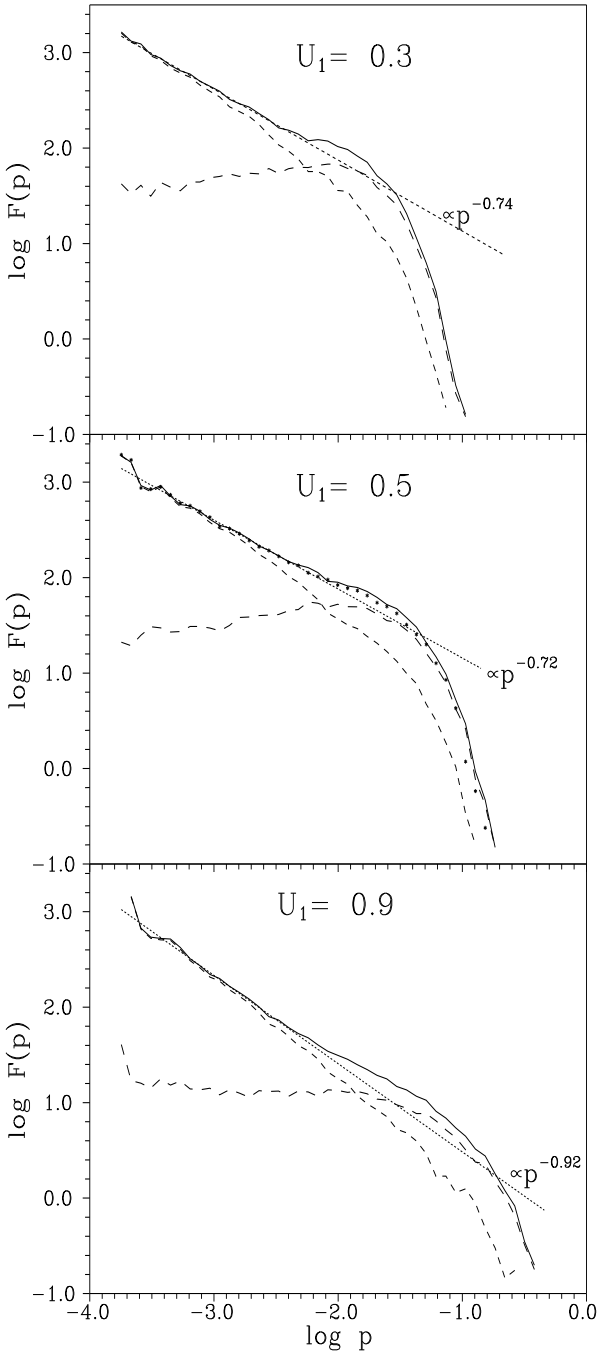}    
\caption[2]{Particle spectra derived for the acceleration process   
involving the first-order Fermi acceleration at the jet terminal shock   
and the acceleration at the jet-cocoon interface. The case with $D =   
0.97$ and a seed particle injection at the shock is considered. The   
spectrum of all escaping particles is presented with a full line, while   
with dashed lines the spectra at the respective escape boundaries:   
medium dashes for the front boundary and long dashes for the side   
boundary. We consider the case with the front boundary placed $L_{esc} =   
R_j$ downstream of the shock and the side boundary in the distance 
$R_{esc} = 2 R_j$ from the jet axis. For the low energy part of the   
front boundary spectrum we provide a power--law fit representing the   
`pure' shock spectrum without the cut-off (short dashes). At the   
successive panels we present the cases for $U_1 = 0.3$, $0.5$ and $0.9$   
. For comparison, the spectrum formed at the shock with neglected energy   
changes at the jet side boundary is presented for the case of $U_1 =   
0.5$ (individual points).} 
\end{figure} 
    
\noindent  
where $\alpha$ is a numerical factor depending on the local conditions 
at, and near the jet, including the escape boundary distance, the 
turbulent magnetic field structure near the flow discontinuity and the 
flow velocity. For the strong scattering limit the estimates of Paper~I 
give the values of $\alpha$ between, say, $10$ and $1$, if we require 
the spectrum to be sufficiently flat, within the velocity range ($0.5$, 
$0.99$) \footnote{One should note that the time scales given in Paper~I 
are derived for stationary spectra. Thus, it provides only the upper 
limits for the actual process, where, at first, the accelerated 
particles appear at the discontinuity and they fill the full diffusive 
volume near the jet in the later times. This filling process is 
accompanied with the gradual flattening of the spectrum. If one inquires 
about the highest particle energies expected to occur at the 
discontinuity, a value of $\alpha$ close to 1 should be assumed in 
Eq.~3. Then, in the energy range close to the upper cut-off, the 
particle spectrum will be steeper than the stationary one (cf. also 
Ostrowski \& Schlickeiser 1996).}. The actual conditions are different 
from the plane-parallel case considered in Paper~I, but we expect the 
above estimates to be still valid, as long as the particle gyroradius 
$r_g(E)$ is smaller than the jet radius $R_j$. If the perturbations are 
due to growing instabilities near the jet boundary, one may expect them 
to reach substantial amplitudes up to and above the kiloparsec scales 
that are of interest here, leading to magnetic field perturbations 
\footnote{An alternative, outside the test particle approach, is 
provided by the anisotropic particle distribution near the boundary, 
inducing the resonance waves' due to streaming instability.} 
$\delta B / B \sim 1$ and $\lambda \sim r_g$ . Then, the acceleration 
time scale becomes 
  
$$T_{acc} = 0.1~\alpha \, \gamma_p \, B_m^{-1} \hskip 5mm [s]    
\hskip 2cm . \eqno(4)$$  
  
For numerical evaluations, let us consider the radio source jet on the   
$\sim 10^{2-3}$ kpc scale of interest here. To estimate the upper energy  
limit for accelerated particles, at first one should compare the time  
scale for energy losses due to radiation and inelastic collisions to 
the acceleration time scale. The discussion of possible loss 
processes is presented by Rachen \& Biermann (1993), who provide the  
loss time scale for protons as  
    
$$T_{loss} \simeq 5\cdot 10^{24}~B_m^{-2} \, (1+Xa)^{-1} \, \gamma_p^{-1}    
 \hskip 0.5cm [s] \eqno(5)$$  
    
\noindent    
where $B_m$ is the magnetic field in mGs units, $a$ is the ratio of the 
energy density of the ambient photon field relative to that of the 
magnetic field, $X$ is a quantity for the relative strength of {\it 
p}$\gamma$ interactions compared to synchrotron radiation and $\gamma_p$ 
is the proton Lorentz factor. For cosmic ray protons the acceleration 
dominates over the losses (Eq-s~4,5) up to the maximum energy 
    
$$\gamma_{p,max} \sim 10^{13} \left[ \alpha \, B_m (1+Xa)  
\right]^{-1/2} \qquad . \eqno(6)$$  
    
\noindent  
This equation can easily yield a large limiting $\gamma_{p,max} \sim 
10^{13}$ with moderate jet parameters (e.g. with $\alpha \approx 10$, 
$B_m \approx 0.1$ and $Xa << 1$). However, one should note that the 
particle gyroradius $r_g(\gamma_p)$ provides the minimum for the 
acceleration region's spatial extent allowing particles to reach the 
predicted value of $\gamma_p$. Thus, for the actual particle maximum 
energy $\gamma^*_{p,max}$ the jet radius should be larger than the 
respective gyroradius $r_g(\gamma^*_{p,max})$ (cf. simulations presented 
in section 3). From the observations of 6 objects Meisenhaimer et al. 
(1989) derived the following `best-guess' parameters for the hot spots 
with emission extending up to the infrared or optical wavelengths: $B_m 
\sim (0.2$ -- $0.8)$ mGs, hot spot diameters $D \sim (0.7$ -- $5.0)$ 
kpc, jet velocities within $(0.1$,  $0.6)$ c and the shock compression 
ratios in the range $(3.5$, $4.8)$.  If, with a bit of optimism, one 
deduces from these estimates the jet (i.e. upstream the shock) 
parameters $B_j \sim 0.2$ mGs, $R_j \sim D \sim 2$ kpc the maximum 
particle energy can reach $10^{20}$ eV ($\gamma_p \sim 10^{10}$). This 
value is substantially smaller than the upper limit given in Eq.~6 and 
scales like $R_j/(2 \, {\rm kpc}) \times B_j/(2 \, {\rm mGs})$. 
    
\subsection{The required efficiency}    
    
Let us consider an isotropic power-law phase-space cosmic ray 
distribution with a cut-off at the momentum $p_{max}$: $f(p) = C \, 
p^{-\sigma} \, H(p_{max} - p)$, where $H$ is the Heaviside step function 
and $\sigma$ is the spectral index. For a flat spectrum, $\sigma < 4.0$, 
the cosmic ray energy density ($\propto$ ${1 \over 4-\sigma} \, 
p_{max}^{4-\sigma}$) peaks near the cut-off. For example, for $\sigma = 
3.0$, 99\% of the cosmic ray energy density falls at the last two energy 
decades. For powerful sources, the rate of energy extraction from the 
galactic nucleus in a form of jet kinetic energy is estimated at $\sim 
10^{59}$ eV/s. If a fraction $\eta$ of this energy is transformed into 
UHE cosmic rays and the particles are transmitted spherically-symmetric 
around the source, then the flux at Earth reaches the value $\sim \eta 
\, 10^7 D_{10}^{-2}$ eV/s/cm$^2$, where $D_{10}$ is the source distance 
in units of $10$ Mpc. Above we neglect losses between the particle 
source and the Earth. As the measurements in the range above 1 EeV give 
the energy flux $\sim 10^4$ eV/cm$^{2}$/s, it is enough to assume a 
small value $\eta$ $\sim$ $10^{-3}$ to explain the observations with a 
single nearby source, or the respectively smaller value for numerous 
sources. For somewhat steeper spectra with $\sigma \approx 4.0$, the 
required particle production efficiency can be an order of magnitude 
larger. The analogous efficiency estimates are obtained by Rachen \& 
Biermann (1993; see also Rachen et al. (1993) and further references 
listed in the first section) and for an alternative 
model, by Kang et al. (1997).

\section{THE ACCELERATED PARTICLE SPECTRUM}    
    
Below, with the use of the Monte Carlo simulations, we discuss   
the spectrum of particles accelerated at the jet if the acceleration   
process at the jet side-boundary is present. Let us note that in   
ultrarelativistic flows both processes - the acceleration at the   
terminal shock and at the considered here tangential discontinuity at   
the jet boundary - are in some way comparable. The mean relative energy   
gain of a particle at individual interaction with the relativistic   
shock, $<\Delta E / E>_{sh} \sim \gamma -1$, where the Lorentz factor  
$\gamma \equiv (1-U^2)^{-1/2}$ and $U = (U_1-U_2)/(1-U_1U_2) \sim 1$.   
An analogous estimate for the boundary acceleration (cf. Eq.~1)  
$<\Delta E / E>_{tang} \sim \gamma_1 -1$ yields a somewhat larger energy 
gain, but for mildly relativistic flows the ratio $(\gamma -1) /  
(\gamma_1 -1) < 1$ is quite close to $1$. Therefore the resulting 
acceleration depends roughly on the number of particle interactions with  
the shock and the boundary discontinuity, respectively. These numbers  
depend on a numerous factors including the injection site of cosmic ray  
particles and the transport properties for these particles. For smaller  
velocities with $\gamma_1 \approx 1$ the shock acceleration becomes the  
first order process with $\Delta E / E \sim U$ and is expected to  
dominate over the second order tangential discontinuity acceleration  
(cf. section 2.2).  
   
As discussed in section (2.1), the spectra of particles accelerated at 
relativistic shock waves depend in a large extent on the detailed   
conditions (magnetic field configuration, amplitude of field   
perturbations, etc.) near the shock. Unfortunately, these conditions   
are usually poorly known and any consideration of the process must be   
based on several rough assumptions. Therefore, in the present simulations 
we do not attempt to reproduce a detailed shape of the particle   
spectrum in any definite astrophysical object, but, rather, we consider   
the form of spectrum modifications introduced to the standard {\it   
power-law with a cut-off} shock spectrum by additional acceleration at   
the jet boundary.  In order to limit the number of free parameters we   
decided to model stationary particle spectra {\it without} taking the   
radiative losses into account, i.e. the upper energy limit of   
accelerated particles (cf. Rachen \& Biermann 1993, Sigl et al. 1995)  
is fixed by the boundary conditions allowing for the escape of the highest 
energy particles. Below, we consider the simplest parallel shock 
configuration. One should note, however, that the derivations of the   
acceleration time scales in relativistic shocks by Naito \& Takahara   
(1995) and Bednarz \& Ostrowski (1996) suggest the possibility of more   
rapid acceleration in shocks with oblique magnetic fields. The   
situation with losses being important is commented in Section 4.   
\begin{figure}  
\vspace {6cm} 
\includegraphics{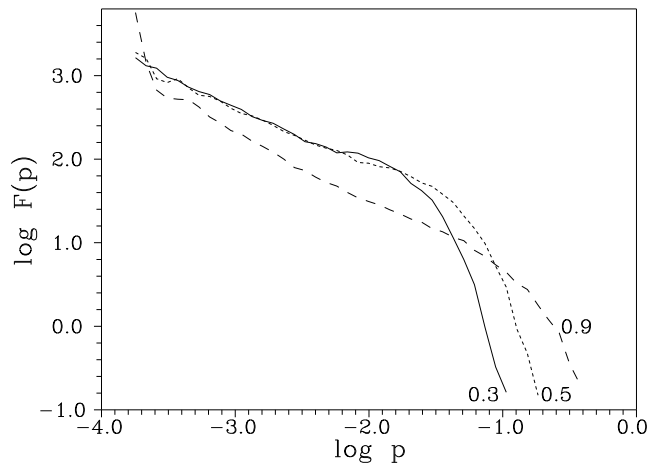}    
\caption[3]{Comparison of the total spectra from Fig.~2 for different    
flow velocities. The velocities are given near the respective curves.}    
\end{figure}    
    
\subsection{Monte Carlo modelling}    
    
In the simulations, we simplify the jet structure near the terminal 
shock as depicted in Fig.~1. The shock is in rest with respect to the 
cocoon surrounding the jet. The upstream plasma hitting the shock moves 
with the relativistic velocity $U_1$ and is advected downstream with the 
velocity, $U_2$. In present simulations we consider three values for the 
flow velocity, $U_1 = 0.3$, $0.5$ and $0.9$ of the light velocity.  The 
shock compression ratio $R \equiv U_1/U_2$ is derived with the use of 
approximate formulae presented by Heavens \& Drury (1988) for shock 
propagating in cold electron--proton plasma and a negligible dynamical 
role of the magnetic field. The analysis of Celotti \& Fabian (1993) 
suggests that ($e$, $p$) plasma can be a viable jet content in the 
strong FRII radio sources. The conditions occurring behind the jet 
terminal shock due to the flow divergence are modelled by imposing the 
particle free escape boundary a finite distance, $L_{esc}$, downstream 
of the shock and in the adjoining front side of the cocoon (`a front 
boundary', cf. Fig.~1). We use the jet radius, $R_j$, as the unit to 
measure $L_{esc}$ and other spatial distances. However, one should note 
that the physical barrier `opposing' particle escape is defined by the 
downstream diffusive scale $L_{diff} \equiv \kappa_2/U_2$, where 
$\kappa_2$ is the downstream diffusion coefficient along the shock 
normal. Here the magnetic field is oriented along this normal and 
$\kappa_2 \equiv \kappa_{\| , 2}$. The value of $L_{esc}$ in units of 
$L_{diff}$ scales as $p^{-1} \, D^{1/2}$, where $D \equiv \kappa_\perp / 
\kappa_\|$ and $\kappa_\|$ ($\kappa_\perp$) is the particle diffusion 
coefficient along (across) the magnetic field. Since the finite extent 
of the actual cocoon and all realistic magnetic field structures therein 
allow for particle escape to the sides we introduce another, tube--like 
free escape boundary surrounding the jet (`a side boundary') in a 
distance of $R_{esc}$ from the jet axis. The unit for the particle 
momentum is defined by the `effective' (cf. Bednarz \& Ostrowski 1996) 
magnetic field, $B_e$, in such a way, that a particle with the momentum 
$p^* = 1.0$ has a gyroradius equal to the jet radius, $r^*_g = p^* c / 
(eB_e) = R_j$. A concept of the effective field is introduced to 
represent the additional magnetic field power contained in waves 
perturbing particle trajectories. The action of the effective field is 
observed in the simulations as bending of the particle trajectory at 
scales smaller than its gyroradius in the uniform component of the 
magnetic field, $B_0$. With a simplified scattering model applied, $B_e$ 
is not uniquely defined (one is unable to evaluate the amount of 
turbulence at scales smaller than $\sim c \, \Delta t$), in the present 
simulations we use the value estimated by Bednarz \& Ostrowski (1996) as 
$B_e = B_0 \sqrt{1 + 4/9 \, (\Delta \Omega / \Delta t )^2}$, where 
$\Delta \Omega$ is the maximum angular momentum scattering amplitude and 
$\Delta t$ is the mean scattering time multiplied by the angular 
gyration velocity (here $= 1$). In the present simulations, the values 
$B_e = 32.0$ and $1.05$ arise for $D = 0.97$ and $0.0013$, the strong 
and the weak scattering cases, respectively. 
   
Below, we consider spectra of particles escaping through the considered 
boundaries for the mono-energetic ($p_0 = 10^{-3}$ or $1.8 \, 10^{-4}$) 
seed particle injection either at the shock ($z_{inj} = 0$) or at the 
jet side boundary far upstream of the shock ($z_{inj} = -10^3 R_j$), and 
for different distances ($L_{esc}$, $R_{esc}$) of the boundaries. The 
particle distribution $F(p) \equiv d\,N(p)\,/\,d({\rm log}\, p)$ 
\footnote{\, \, For the power-law distribution the spectral index for 
$F(p)$ is $\sigma - 3$ (cf. Sect.~2.4).}, which gives the particle 
number per logarithmic momentum bandwidth, is derived for particles 
escaping through the boundaries. For simplicity, in order to limit the 
number of free parameters in the simulations, we assume the mean 
magnetic field to be parallel to the jet velocity both within the jet 
and in the cocoon (cf. Coleman \& Bicknell 1988, Fedorenko \& 
Courvoisier 1996). In the examples presented below we consider a 
(resulting from simulations) ratio $D \equiv \kappa_\perp / \kappa_\|$ 
to be either $0.0013$ ({\it small}\,) or $0.97$ ({\it large} cross field 
diffusion). We consider these values as the effective ones, representing 
the transport properties of the medium with realistic magnetic field 
structures. The geometric pattern of the introduced particle trajectory 
perturbations' is taken to be the same at all momenta (Ostrowski 1991). 
Thus both diffusion coefficients are proportional to the particle momentum 
and the above ratio of diffusion coefficients is constant. Further 
details of the simulations are described in Appendix A. 
\begin{figure}    
\vspace{6cm} 
\includegraphics{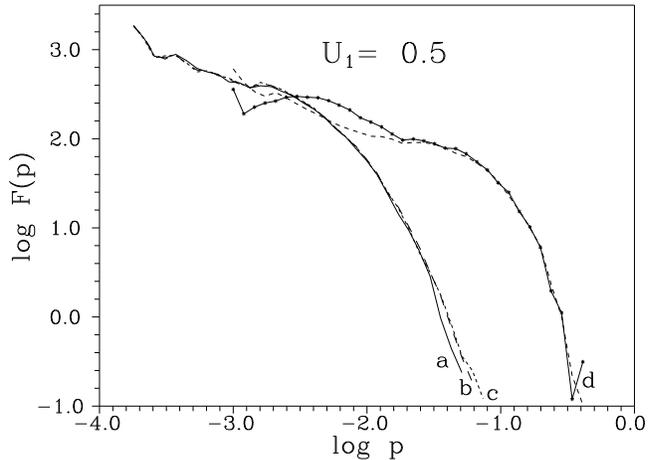}    
\caption[4]{Comparison of particle spectra for different distances to the  
jet side boundary for $U_1 = 0.5$. The case with $D = 0.97$ and the seed  
particle injection at the shock is considered. Three spectra are  
presented for $L_{esc} = 1.0$: a.) $R_{esc} = 1.1$, b.) $R_{esc} = 2.0$,  
c.) $R_{esc} = 11.0$, and two additional spectra for $R_{esc} = 11.0$:  
d.) $L_{esc} = 10.0$ and e.) $L_{esc} = 100.0$.} 
\end{figure}  
   
At Fig-s~2-4, we consider particle spectra for the seed particle 
injection at the shock. The fixed spatial distances to the escape 
boundaries are assumed, but the size of particle trajectory defined by 
its gyroradius, as well as the spatial diffusion coefficients, increase 
in proportion to the particle momentum. As a result, the escape 
probability grows with particle energy providing a cut-off in the 
spectrum. The energy scale of the cut-off is different for the front 
boundary spectrum and the side boundary spectrum, with the latter being 
often larger in our simulations. This difference can occur also in the 
case of numerical experiments with the jet boundary acceleration turned 
off. It is due to the fact that particles accelerated at the shock close 
to the jet boundary have the opportunity to diffuse back upstream across 
the static cocoon medium, to be accelerated again at the shock and then 
escape through the side boundary. The difference between these two 
scales increases for example, with the jet velocity, the extent of the 
diffusive cocoon, shifting the particle injection site upstream of the 
shock, increasing the effective particle radial diffusion coefficient. 
However, one should note that the role of the jet boundary acceleration 
is limited for the seed particle injection at the shock in the presence 
of the nearby front escape boundary (cf. Fig.~2, middle panel). As 
explained below the situation will change drastically for the injection 
at the jet boundary far upstream of the shock. 
   
In the spectra presented at Fig.~2 three parts can be clearly separated. 
The first one, with a wavy behaviour, reflects the initial conditions of 
the mono-energetic injected spectrum interacting with the accelerating 
surfaces -- for the shock injection only the shock acceleration is 
important in this range. In the remaining part of the spectrum, at 
energies directly preceding the cut-off energy, the spectrum exhibits 
some flattening with respect to the inclination of the lower energy 
part. The low energy section of the spectrum -- within computational 
accuracy -- coincides with the analytically derived inclination of the 
spectrum formed at the infinitely extended shock (cf. Heavens \& Drury 
1988). The spectrum flattening at larger energies occurs due to 
additional particle transport from the shock's downstream region to the 
upstream one through the cocoon surrounding the jet (this effect occurs 
also if there is no side boundary acceleration ! ), and inclusion of a 
very flat spectral component resulting from the side boundary 
acceleration (see below). 
   
A comparison of particle spectra generated at jets with different 
velocities is presented in Fig.~3. One may note a systematic shift of 
the spectrum cut-off toward higher energies with an increase of the jet 
velocity. Additionally, at the low energy portion of the spectrum, the 
expected spectral index change can be observed. The influence of varying 
distances ($L_{esc}$, $R_{esc}$) at particle spectra can be evaluated by 
inspecting Fig.~4.  Decreasing any boundary distance leads to decreasing 
the cut-off energy, however the actual changes depend in a substantial 
degree on the leading escape process removing particles from the 
acceleration region - either the diffusive/free-escape through the front 
boundary or the radial diffusion toward the side boundary. Let us 
finally stress, that even for an infinite diffusive volume surrounding the 
jet, the acceleration efficiency will decrease for particles with momenta 
$p > p^*$~. Therefore, the upper momentum cut-off can not reach values 
above the scale $ \sim \gamma_u p^* $~. 
    
\begin{figure} 
\vspace{6cm} 
\includegraphics{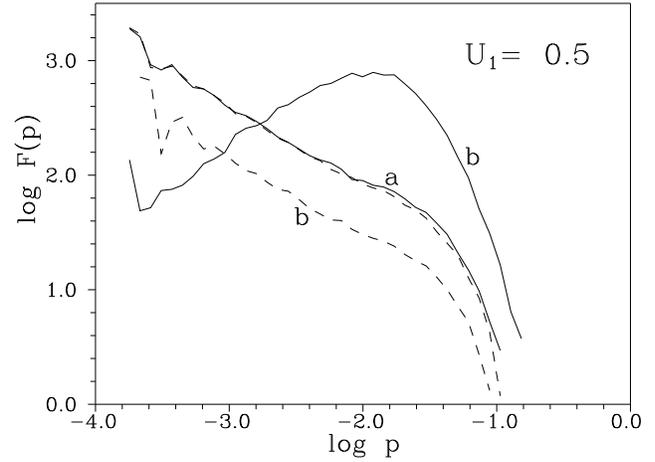}    
\caption[5]{The particle spectra formed with and without the jet boundary   
acceleration for $D = 0.97 $. Spectra formed due to acceleration both at   
the jet boundary and at the terminal shock are presented with full   
lines, while the spectra for the neglected boundary acceleration are 
given with dashed lines. The results are presented for $R_{esc} = 2.0$,   
$L_{esc} = 1.0$ and a.) $z_{inj} = 0.0$ or b.) $z_{inj} = -1000.0$.}   
\end{figure}   
    
\begin{figure} 
\vspace{6cm} 
\includegraphics{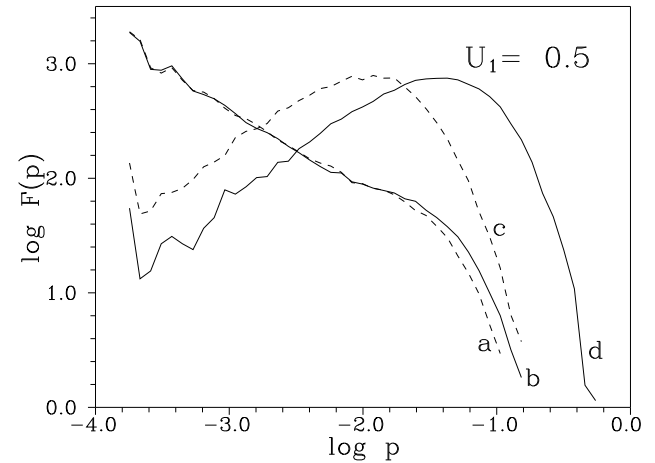}    
\caption[6]{ Comparison of the particle spectra formed with wide   
($R_{esc} = 11.0$; dashed lines (a) and (c) ) and narrow ($R_{esc} =   
2.0$; full lines (b) and (d) ) cocoon. The results are presented for   
$L_{esc} = 1.0$ and two possibilities for particle injection: $z_{inj} =   
0.0$ for cases (a) and (b), and $z_{inj} = -1000.0$ for the cases (c)   
and (d). }  
\end{figure}

In Fig.~5 we compare spectra of particles injected at the jet side 
boundary far upstream of the shock, in the distance $10^3 \, R_j$, to 
the spectra of particles injected at the terminal shock. In the former 
case resulting distributions are very flat or even inverted ($\sigma < 
3$, cf. Appendix B). This feature results from the character of the 
acceleration process with particles having an opportunity to hit the 
accelerating surface again and again due to an inefficient diffusive 
escape to the sides. The apparent deficiency of low-energy particles in 
these spectra results from the fact that most of these particles 
succeeded in crossing the discontinuity several times before they are 
able to diffusively escape through the side boundary. In other words, 
the escape due to particle energy increase (and the corresponding 
diffusion coefficient increase) is much  more effective than the escape 
caused by diffusion of the low energy particles across the cocoon. If we 
neglect the tangential discontinuity acceleration, the particle advection 
toward the terminal shock appended with the side boundary diffusive 
escape determines an initial phase of the upstream transport. The 
final energy spectrum is formed at the shock. 
   
We also performed simulations with the continuous injection process  
extended along the jet boundary from $z = -1000$ up to the shock at $z =  
0$. As long as most particles were injected far from the shock the 
resulting spectra were very similar to the ones obtained with $z_{inj} =  
-1000$.  
   
The role of the cocoon spatial extension in the acceleration process can 
be evaluated from Fig-s~4,6. One may note that for the particle 
injection at the terminal shock in the presence of a nearby front 
boundary, any change in $R_{esc}$ is accompanied by only a minor 
modification to the spectrum seen at highest energies. This behaviour is 
determined by the efficient particle escape through the front boundary. 
Only increasing $L_{esc}$ to values $\gg L_{diff}(p=p_0)$ allows the 
larger $R_{esc}$ to increase the spectrum cut-off substantially. 
However, for the injection far upstream of the shock the main process 
removing particles from the acceleration is the diffusive escape through 
the side boundary.  In such a case particles can reach larger energies 
with a more extended cocoon. For the shock injection, the spectra 
presented in Fig.~6 only insignificantly differ in the momentum range 
between $10^{-4}$ and $10^{-2}$. This feature illustrates the fact that 
spectrum inclination depends on the local conditions near the shock if 
the particle energy is insufficient to allow for non-diffusive escape 
through the boundary. For the upstream injection the spectrum with 
smaller $R_{esc}$ is shifted up in this momentum range because the 
larger proportion of all particles have a chance to escape at a given 
momentum. However, the spectrum inclination does not significantly 
depend on $R_{esc}$ until a chance for particle escape becomes 
comparable to the probability of doubling its energy. The process can be 
described in the following way. After commencing the acceleration 
process at injection, the energized particles fill diffusively the 
volume near $R_j$. The normal to the jet boundary diffusion coefficient 
is proportional to particle momentum in our model. Thus the diffusing 
relativistic particles fill this volume in a time inversely proportional 
to the particle momentum, $\propto p^{-1}$, and the time required for 
these particles to diffuse back to the jet boundary to be further 
accelerated is also $\propto p^{-1}$. As energy gains of particles 
interacting with the jet boundary - with the mean value depending on the 
particle anisotropy and the jet velocity - are proportional to $p$, the 
resulting spectrum inclination only slightly depends on the size 
$R_{esc}$ at energies much smaller than the cut-off energy. A simplified 
analytic approach to this acceleration process is presented in Appendix 
B. 
   
In Fig.~7, the spectra for different turbulence levels defined by the 
respective values of $D = 0.97$ or $0.0013$ are compared. Because in the 
simulations with smaller $D$, particles were injected at larger initial 
momentum, in order to make the comparison more simple, we scaled (i.e. 
vertically shifted) the spectrum (a) to coincide in the power-law 
section with the spectrum (b). One can observe that the cross-field 
diffusion, changing as $\approx D^{1/2}$, has a substantial influence on 
the spectrum if the particle radial diffusion is the main process 
removing particles from the jet vicinity. 
   
\begin{figure} 
\vspace{6cm} 
\includegraphics{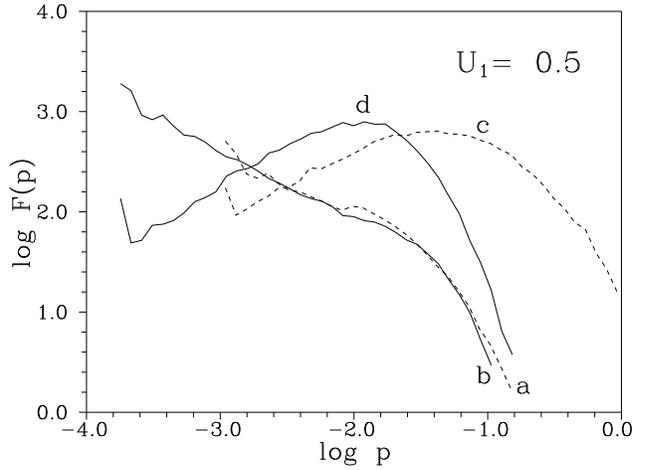}    
\caption[7]{Particle spectra formed with $D = 0.97 $ (full lines; $p_0 =   
1.8\,10^{-4}$) versus the spectra for the small $D = 0.0013$ (dashed   
lines; $p_0 = 10^{-3}$). The spectra (a) and (b) are formed for the   
shock injection $z_{inj} = 0$, while the spectra (c) and (d) for the jet   
boundary injection at $z_{inj} = -1000.0$. All presented results were   
derived with $R_{esc} = 2.0$ and $L_{esc} = 1.0$.} 
\end{figure}   
   
\section{FINAL REMARKS}    
    
Application of the `pure' shock acceleration mechanism in order to model 
UHE particle production in the discussed extragalactic jets meets a 
number of difficulties. Due to small spatial dimensions of hot spot 
regions, the diffusive particle escape may become a serious problem for 
particles, preventing them from reaching energies above $10^{20}$ eV. We 
would like to note that the respective losses would be enhanced in the 
presence of oblique fields causing the drift motions along the shock 
surface. Therefore it is not clear if the speeding up of the 
acceleration process at oblique shocks will be accompanied by the 
spectrum cut-off energy increase. In the super-luminal magnetic field 
configuration, the acceleration to UHE can be completely suppressed if 
there is no other mechanism providing particles in the sub-UHE energy 
range (cf. Begelman \& Kirk 1990, Ostrowski 1993a). Another difficulty 
may arise due to the high inclination of the spectrum, $\sigma > 4.0$, 
expected in some cases in relativistic shocks (Ostrowski 1991, 1993a; 
Ballard \& Heavens 1993). In realistic conditions it can substantially 
increase  the shock wave energy conversion into lower energy cosmic 
rays. The total efficiency of cosmic ray generation must then be much 
larger to allow for the required power in UHE particles. 
   
The presented hypothesis for the acceleration mechanism producing the 
most energetic cosmic ray particles at the jet side-boundary has a 
number of advantages over the `pure' shock hypothesis. At first, its 
efficient action is rather weakly dependent on detailed conditions in 
the acceleration site. The mean magnetic field elongated along the jet 
axis assumed in the present paper is in no way essential to the 
presented discussion. One may note, however, that such a field 
configuration can be produced in a natural way due to any form of 
viscosity near the jet boundary, including the viscous pull of 
magnetized plasma mediated by the accelerated particles (Eq.~2, cf. 
Paper~I). Then, the field perturbations required near the boundary are 
created by growing short wave instabilities at the jet boundary surface, 
as well as by the anisotropically distributed cosmic rays streaming 
along the boundary.  The required energy expense of the jet for the UHE 
cosmic ray production is quite reasonable in comparison to the energy 
available. Within the present model involving relativistic jets, as long 
as global instabilities do not disrupt the organized jet flow, the 
cut-off energy can be a factor of a few ones greater than the value 
obtained in the pure shock acceleration if seed particles for 
acceleration are injected all the way along the jet boundary. Then, the 
obtained spectrum can be very flat before the cut-off. One may note that 
no compression or de-compression is directly connected with this 
acceleration process and adiabatic losses seem not to be an obstacle 
here. A slow perpendicular expansion of the jet is in most cases 
controlled by the magnetic field structure and we do not expect it to 
play any noticeable role in particle deceleration within the UHE energy 
range. To recapitulate, we believe that because the respective 
conditions arise in a natural way, the process of particle acceleration 
at tangential discontinuity should be seriously considered as an 
important supplementary process to shock acceleration. 
   
The main difficulty in analysing the details of the particle spectrum 
comes from an inadequate knowledge of local conditions in the   
acceleration region inside the jet and in the space surrounding the   
jet, including the magnetic field strength and configuration, the form   
and the amplitude of field perturbations, the distance of the   
`effective' escape boundary or, finally, at low particle energies, the   
velocity profile of the turbulent shear layer expected to occur at the   
interface between the jet and the surrounding medium. Of course,   
observations provide some information about magnetic field magnitude   
and the mean field structure, but, besides its fragmentary, projection   
and resolution dependent character, the measurements relate to regions   
where radiative losses of cosmic ray electrons take place, not   
necessarily strictly coinciding with the ion acceleration sites. The   
information to be borrowed from hydrodynamical jet simulations usually   
refers to the flows with limited values of the magnetic field (cf. Marti   
et al. 1995, 1997; G\'omez et al. 1995). So, the available basic information  
required for the cosmic ray spectrum derivation is very limited.   
Fortunately, the rapid acceleration at the velocity discontinuity   
results in very flat spectra in all situations where particles near the   
discontinuity have a chance to cross this surface at least few times   
before the escape. When considering the energy budget of accelerated  
particles, the flat spectrum means that cosmic ray energy is contained   
in particles with highest energies. In the conditions of effective   
acceleration, the upper scale for the UHE particle energy density is 
provided by the magnetic field energy density near the jet boundary.   
Possessing the higher energy density, UHE cosmic rays would smear out the   
discontinuity into a wide shear layer with a much reduced acceleration 
efficiency.   
   
Finally, let us comment on the particle spectrum in the presence of   
radiative loses decreasing the cut-off energy in the spectrum. If the   
loss process can be sufficiently effective to shift the cut-off energy   
substantially below the `geometric' cut-off present in our simulations,   
there will occur weaker mixing of the two acceleration processes   
in forming the total spectrum. No mixing means that there are two   
acceleration time scales present, the one for the shock acceleration   
and the other one for the tangential discontinuity acceleration. The   
spectra formed in these two processes -- at the shock and at the jet  
boundary far from the shock -- can be independent, with a different  
shape and an energy cut-off. For electrons the radiative cut-off can  
occur at such low energies that the acceleration process at the jet 
boundary involves only other processes (e.g. the second-order Fermi  
acceleration, highly oblique shocks, magnetic field reconnection) in the  
turbulent boundary layer, with the viscous shear acceleration playing  
only a secondary role (cf. Ostrowski 1997). Without a detailed 
consideration it is difficult to draw any conclusions about the resulting 
synchrotron spectra. We would like to note, however, that some  
observations of the synchrotron optical jets may require such mechanisms to 
operate (Ostrowski, in preparation). Till now the most detailed  
information available about the synchrotron jet structure is for the 
nearby M87 jet (e.g. Sparks et al. 1996, and references therein), but at  
least five more have been observed. However, the observations are not  
conclusive in the matter of the particular mechanism responsible for the 
energetic particle populations present in these sources.  
    
\acknowledgements{The present work was supported through the grants PB   
1117/P3/94/06 and PB 179/P03/96/11 from the {\it Komitet Bada\'n   
Nauko\-wych}.}   
   
\section*{Appendix~A: Modelling of the particle acceleration process}  
    
We consider the particle diffusion in the jet-cocoon region in the case of 
the mean magnetic field being parallel to the jet velocity both inside   
and outside the jet. Modelling of the diffusive particle trajectories is   
based on the small-amplitude pitch-angle scattering approach (Ostrowski   
1991). It assumes that the particle diffusion coefficients parallel to   
the mean field, $\kappa_\|$, and the one perpendicular to the mean   
field, $\kappa_\perp$, are proportional to the particle momentum. Due to 
this fact, the amount of computations required for reproducing particle 
diffusive trajectories for low energy particles is much larger than the   
respective amount for particles with higher energies. Therefore, in 
order to deal with these low energy particles, we introduced a hybrid 
approach to derive particle trajectories. For particles near the 
surfaces of the flow velocity change at the jet boundary and at the 
shock, and not further than $2$ particle gyroradii away in the former 
case and $1$ diffusive scale $\kappa_{\parallel,i} / U_i$ ($i = 1$, $2$) 
from the shock surface, we use the exact form of the mentioned 
pitch-angle scattering method. When the particle diffuses further away 
from the respective discontinuity, we use the spatial diffusion model 
involving large time steps. However, the step length is always limited 
in order to prevent particles from crossing any discontinuity in an 
individual `diffusive' step - the mean diffusive step allows only for 
moving a particle $0.25$ of the distance to its closest boundary. In 
these computations we use the values of $\kappa_\|$ and $\kappa_\perp$ 
derived in independent simulations involving the `exact' pitch-angle 
diffusion procedure. This way we are able to consider particles with 
momenta which differ in a few orders of magnitude. During the 
simulations we use a variant of the trajectory splitting procedure 
described by Ostrowski (1991). A good test for this simulation procedure 
is provided by the agreement of our shock accelerated spectra fitted at 
low energies with the values derived analytically by Heavens \& Drury 
(1988). 
  
\section*{Appendix~B: On the spectrum of particles accelerated at the  
tangential discontinuity}  
  
Let us consider a simple model for particles accelerated at the plane 
tangential discontinuity surrounded with infinite regions for particle 
diffusion. We derive the stationary spectrum at the discontinuity for 
the perpendicular (to discontinuity) diffusion coefficient being 
proportional to particle energy, $\kappa_\perp \propto E$. For 
relativistic particles with $p = E$ and the same velocity $v = c$, the 
mean time between the successive particle interactions with the 
discontinuity is also proportional to the particle's energy. Thus the 
mean rate of particle energy gain is constant, independent of energy, 
$<\dot{p}>$ = const. The transport equation for the {\it phase-space} 
distribution function $f(p)$ has the following form, 
  
$$ {1 \over p^2} {\partial \over \partial p} \left[ p^2 <\dot{p}> f(p)  
\right] = Q_0 \delta (p-p_0) \qquad . \eqno(B1)$$  
  
\noindent  
In the above equation one assumes a continuous particle injection at $p 
= p_0$. For a general power-law form for $<\dot{p}> = C_0 p^\alpha$ it 
yields the solution, 
  
$$f(p) = {Q_0 p_0^2 \over C_0} \, p^{-2-\alpha} \, H(p-p_0)  \qquad ,  
\eqno(B2)$$  
  
\noindent  
where $H(x)$ is the Heaviside step function. As long as there are no  
specific energy scales introduced into the acceleration process the  
obtained form for $f(p)$ is independent of the velocity difference at  
the discontinuity (particle anisotropy), and of possible correlation  
between the interaction time and the energy gain. In the case considered  
in the present paper $\alpha = 0$ and $f(p) \propto p^{-2}$.  
  
For the discontinuity formed at the jet boundary, the jet radius and the 
escape boundary radius provide energy scales to the process. As a 
result, a cut-off occurs at large energies in the spectrum. However, at 
small energies, where the jet boundary curvature is insignificant and 
the diffusive regions very extended, the solution should be close to the 
one given in (B2). In fact, numerical fits for low energy sections of 
cases (c) and (d) at Fig.~6 give the respective fits $f(p) \propto 
p^{-2.16}$ and $\propto p^{-2.10}$ (the difference in the obtained 
spectral indices is not significant, depending on the momentum range 
used for fitting).

\section*{References}    
 Ballard K.R., Heavens A.F., 1991, MNRAS, {\bf 251}, 438   \\ 
 Ballard K.R., Heavens A.F., 1992, MNRAS, {\bf 259}, 89   \\ 
 Begelman M.C., Kirk J.G., 1990, ApJ, {\bf 353}, 66   \\ 
 Bednarz J., Ostrowski M., 1996, MNRAS, {\bf 283}, 447 \\ 
 Berezhko E.G., 1990, Preprint {\it Frictional Acceleration of  \par 
Cosmic Rays}, The Yakut Scientific Centre, Yakutsk.   \\ 
 Biermann P.L., 1994, Lecture at the Chinese Academy of  \par Sciences -  
Max Planck Society Joint Seminar, Nandaihe, \par China (MPIfR preprint).   \\ 
 Biermann P.L., Rachen J.P., Stanev T., 1995, in Proc. 24th  
Int.  \par Cosmic Ray Conf., Rome \\ 
 Bridle A.H., Hough D.H., Lonsdale C.J., Burns J.O., Laing  \par 
R.A., 1994, AJ, 108, 766 \\ 
 Celotti A., Fabian A.C., 1993, MNRAS, {\bf 264}, 228 \\ 
 Coleman C.S., Bicknell G.V., 1988, MNRAS, {\bf 230}, 497 \\ 
 Earl J.A., Jokipii J.R., Morfill G., 1988, ApJ, {\bf 331}, L91 \\ 
 Elbert J.W., Sommers P., 1995, ApJ, {\bf 441}, 151 \\ 
 Ellison D.C., Jones F.C., Reynolds S.P., 1990, ApJ, {\bf 360}, 702   \\ 
 Fedorenko V.N., Courvoisier T.J.-L., 1996, A{\&}A, {\bf 307}, 347 \\ 
 Geddes J., Quinn C., Wald R.M., 1996, ApJ, {\bf 459}, 384. \\ 
 Ghisellini G., Padovani P., Celotti A., Maraschi L., 1996,  \par 
ApJ, {\bf 407}, 65                                                \\ 
 G\'omez J.L., Marti J.M$^{\underline{\rm a}}$, Marscher A.P.,  
Ib\'a\~nez J.M$^{\underline{\rm a}}$,  \par Marcaide J.M., 1995, ApJ, {\bf 449},  
L19  \\ 
 Hayashida et al., 1996, Phys. Rev. Lett., {\bf 77}, 1000. \\ 
 Heavens A., Drury L'O.C., 1988, MNRAS, {\bf 235}, 997.   \\ 
 Henriksen R.N., 1987, ApJ, {\bf 314}, 33. \\ 
 Jokipii J.R., Kota J., Morfill G., 1989, ApJ, {\bf 345}, L67. \\ 
 Kang H., Ryu D., Jones T.W., 1996, ApJ, {\bf 456}, 422. \\ 
 Kang H., Rachen J.P., Biermann P.L., 1997, MNRAS, {\bf 286}, \par  257. \\ 
 Kirk J.G., 1997, in {\it Relativistic Jets in AGNs}, eds.   M.  
Os- \par trowski, M. Sikora, G. Madejski, M. Begelman (Cracow) \\ 
 Kirk J.G., Heavens A., 1989, MNRAS, {\bf 239}, 995   \\ 
 Kirk J.G., Schneider P., 1987, ApJ, {\bf 315}, 425   \\ 
 Kirk J.G., Schneider P., 1988, A\&A, {\bf 201}, 177   \\ 
 Marti J.M$^{\underline{\rm a}}$, M\"uller E., Font J.A.,  
Ib\'a\~nez J.M$^{\underline{\rm a}}$, 1995, ApJ, {\bf 448},  \par L105 \\ 
 Marti J.M$^{\underline{\rm a}}$, M\"uller E., Font J.A.,  
Ib\'a\~nez J.M$^{\underline{\rm a}}$, Marquina A.,  \par 1997, ApJ, {\bf  
479}, 151.  \\ 
 Medina Tanco G.A., 1998, ApJL (in press).\\ 
 Medina Tanco G.A., de Gouveia Dal Pino E.M., Horvath J.E.,  \par  1997,  
Astropart. Phys., {\bf 6}, 337.  \\ 
 Meisenheimer K., R\"oser H.-J., Hiltner P.R., Yates M.G.,  
Lon-  \par gair M.S., Chini R., Perley R.A., 1989, A{\&}A, {\bf 219}, 63 \\ 
 Naito T., Takahara F., 1995, MNRAS, {\bf 275}, 1077   \\ 
 Newman P.L., Moussas X., Quenby J.J., Valdes-Galicia J.F.,   \par   
 Theodossiou-Ekaterinidi Z., 1992, A\&A, {\bf 255}, 443.  \\ 
 Ostrowski M., 1990, A\&A, {\bf 238}, 435 (Paper~I). \\ 
 Ostrowski M., 1991, MNRAS, {\bf 249 }, 551. \\ 
 Ostrowski M., 1993a, MNRAS, {\bf 264}, 248. \\ 
 Ostrowski M., 1993b, in Proc. 23rd Int. Cosmic Ray  
Conf., p.  \par 329 (OG 9.3.6) Calgary   \\ 
 Ostrowski M., 1994, Comments on Astrophys., 17, 207. \\ 
 Ostrowski M., 1996, in Proc. ICRR Symp. on {\it Highest Energy \par Cosmic  
Rays}, ed. M. Nagano, Tanashi (p. 377) \\ 
 Ostrowski M., 1997, in {\it Relativistic Jets in AGNs}, eds. M.  
Os-  \par trowski, M. Sikora, G. Madejski, M. Begelman (Cracow) \\ 
 Ostrowski M., Schlickeiser R., 1996, Solar Phys., {\bf 167}, 381. \\ 
 Rachen J.P., Biermann P., 1993, A\&A, {\bf 272}, 161. \\ 
 Rachen J.P., Stanev T., Biermann P., 1993, A\&A, {\bf 273}, 377. \\ 
Sigl G., 1996, Space Sci. Rev., {\bf 75}, 375. \\ 
 Sigl G., Schramm D.N., Bhattacharjee P., 1995, Astropart.  \par 
Phys. {\bf 2}, 401.   \\ 
 Sigl G., Schramm D.N., Lee S., Coppi P., Hill Ch.T., 1996, ApJ  \par 
Lett. (submitted, ASTRO-PH/9605158).   \\ 
 Sparks W.B., Biretta J.A., Macchetto F., 1996, ApJ, {\bf 473}, 254 \\ 
 Stanev T., Biermann P.L., Lloyd-Evans J., Rachen J.P.,  \par Watson  
A.A., 1995, Phys. Rev. Lett., {\bf 75}, 3056.   \\ 
\end{document}